\documentclass[life,article,accept,pdftex,moreauthors]{Definitions/mdpi} 

\firstpage{1} 
\pubvolume{1}
\issuenum{1}
\articlenumber{0}
\pubyear{2025}
\copyrightyear{2025}
\externaleditor{Amedeo Balbi}
\datereceived{10 December 2024} 
\daterevised{6 January 2025} % Comment out if no revised date
\dateaccepted{ } 
\datepublished{ } 
\hreflink{https://doi.org/} % If needed use \linebreak
\usepackage{soul} % For highlighting

\Title{{The} Role of Atmospheric Composition in Defining the Habitable Zone Limits and Supporting \textit{E. coli} Growth}

\TitleCitation{The Role of Atmospheric Composition in Defining the Habitable Zone Limits and Supporting \textit{E. coli} Growth}

\Author{{Asena Kuzucan} $^{1,2,}$*\orcidA{}, Emeline Bolmont $^{1,2}$\orcidB{}, Guillaume Chaverot $^{2,3}$\orcidC{}, Jaqueline {Quirino} Ferreira $^{2,4}$\orcidD{}, \mbox{Bastiaan {Willem} Ibelings $^{2,4}$\orcidE{},}  Siddharth Bhatnagar $^{1,2,5}$\orcidF{} and Daniel {Frank} McGinnis $^{2,4}$\orcidG{}} 

\AuthorNames{Firstname Lastname, Firstname Lastname and Firstname Lastname}

\AuthorCitation{Kuzucan, A.; Bolmont, E.; Chaverot, G.; Ferreira, J.Q.; Ibelings, B.W.; Bhatnagar, S.; McGinnis, D.F.}
\address{%
$^{1}$ \quad Observatoire de Gen\`eve, Universit\'e de Gen\`eve, Chemin Pegasi 51, 1290 Versoix, Switzerland; {emeline.bolmont@unige.ch}  (E.B.); {siddharth.bhatnagar@unige.ch} (S.B.) \\ 

$^{2}$ \quad Centre sur la Vie dans l'Univers, Universit\'e de Gen\`eve, {1211 Geneva, Switzerland}; {guillaume.chaverot@univ-grenoble-alpes.fr} (G.C.); {jaqueline.quirinoferreira@unige.ch} (J.Q.F.); {bastiaan.ibelings@unige.ch} (B.W.I.); {daniel.mcginnis@unige.ch} (D.F.M.)\\

$^{3}$ \quad CNRS, IPAG, {University} Grenoble Alpes,  F-38000 Grenoble, France\\
 
$^{4}$ \quad Department F.-A. FOREL for Environmental and Aquatic Sciences, Universit\'e de Gen\`eve, {1211 Gen\`eve, Switzerland}\\

$^{5}$ \quad Department of Applied Physics {and} Institute for Environmental Sciences, Universit\'e de Gen\`eve, {1211 Gen\`eve, Switzerland}} 

\corres{Correspondence: asena.kuzucan@unige.ch or {asenakuzucan@gmail.com} } 

\abstract{Studying exoplanet atmospheres is essential for assessing their potential to host liquid water and their capacity to support life (their habitability). Each atmosphere uniquely influences the likelihood of surface liquid water, defining the habitable zone (HZ)—the region around a star where liquid water can exist. However, being within the HZ does not guarantee habitability, as life requires more than just liquid water. In this study, we adopted a two-pronged approach. First, we estimated the surface conditions of planets near the HZ's inner edge under various atmospheric compositions. By utilizing a 3D climate model, we refined the inner boundaries of the HZ for planets with atmospheres dominated by H$_2$ and CO$_2$ for the first time. Second, we investigated microbial survival in these environments, conducting laboratory experiments on the growth and survival of \textit{E. coli} K-12, focusing on the impact of different gas compositions. This innovative combination of climate modeling and biological experiments bridges theoretical climate predictions with biological outcomes. Our findings indicate that atmospheric composition significantly affects bacterial growth patterns, highlighting the importance of considering diverse atmospheres in evaluating exoplanet habitability and advancing the search for life beyond Earth.}


\begin{document}

%%%%%%%%%%%%%%%%%%%%%%%%%%%%%%%%%%%%%%%%%%
%\endnote{This is an endnote.} % To use endnotes, please un-comment \printendnotes below (before References). Only journal Laws uses \footnote.

% The order of the section titles is different for some journals. Please refer to the "Instructions for Authors” on the journal homepage.

\section{Introduction}\label{introduction}

The search for life beyond our planet has been stimulated by the identification of more than 5765 confirmed exoplanets within 4304 planetary systems, as of 4 October  2024~\citep{NASA}. Some of these celestial bodies, such as the TRAPPIST-1 planets \citep{Gillon_2017, Turbet_2020a} and Proxima-b \citep{Anglada_2016, Turbet2016}, may offer potential habitats for life. Given the adaptability of life to diverse and extreme conditions on Earth \citep{Horikoshi}, it is plausible that similar environments on exoplanets could support life.

Life as we know it requires several key factors to thrive, including appropriate surface conditions, an energy source (such as a star or geothermal heat), essential elements such as carbon, hydrogen, nitrogen, oxygen, phosphorus, and sulfur (CHNOPS), other elements specific to organisms (e.g., Na, K, and Cu), and most importantly, liquid water \citep{Cockell}. Water is essential for metabolic activities, regardless of environmental conditions. Therefore, in the search for extraterrestrial life, particular attention should be paid to planets within the HZ, where conditions allow for the presence of liquid water on the surface \citep{Kasting93, Kopparapu}. However, the HZ specifically refers to the potential for surface liquid water to be present, which is only one of several prerequisites for life. The extent of the HZ varies depending on factors like the type of host star, planet size, and atmospheric composition \citep{Cockell}. 

The presence of liquid water within the HZ does not necessarily imply that an environment is ``habitable'' for specific life forms. In this study, habitability refers to the environmental conditions necessary for the survival and growth of \textit{E. coli} and other Earth-like microorganisms, which depend not only on the availability of liquid water but also on suitable atmospheric compositions and surface conditions. For example, a planet within the HZ may contain liquid water, but its atmosphere might not support metabolic needs or provide adequate protection from harmful radiation, both of which are critical for sustaining life as we know it.

As observational characterization of the atmospheres of small temperate planets is still ongoing (\citep{Greene, Zieba_2023} for TRAPPIST-1 b and c), it is crucial to study the HZ limits for different atmospheric compositions \citep{Chaverot_2023}, using atmospheric modeling. The inner edge of the HZ has been extensively studied using 1D models, but less so with 3D models. Additionally, previous 3D studies have typically considered only a narrow range of gases, often focusing on Earth-like atmospheres with slight variations \citep{Leconte_2013,Kopparapu_2016, Kopparapu_2017, Way2020, Chaverot_2023, Turbet_2023}. This study is the first to explore the limits of the HZ in 3D for two compositions that have not been previously investigated (H$_2$- and CO$_2$-dominated), while also investigating whether these conditions can sustain life, specifically focusing on microbial adaptability.

Terrestrial planets can develop their atmospheres through the capture of nebular gases, degassing during the accretion process, or subsequent degassing resulting from volcanic activity \citep{Elkins-Tanton_2008, SCHAEFER2010}. Based on these processes and the nature of the interior of the planet, Hu et al. (2012) \cite{Hu_2012} described the standard cases for terrestrial atmospheres as being reducing (H$_2$-rich), weakly oxidizing (N$_2$-rich), or oxidizing (CO$_2$-rich). Based on these benchmarks, we focus on the same three atmospheric compositions. The HZ limits for N$_2$-dominated atmospheres have been extensively studied using both 1D climate models \citep{Kopparapu, Ramirez_2020}, and 3D Global Climate Models (GCMs) for fast and slow rotators \citep{Kopparapu_2017, Leconte2013, Wolf2015, Chaverot_2023}. Our study includes simulations using an N$_2$ + 376 ppm CO$_2$ atmosphere for validation purposes, i.e., compared with \citep{Chaverot_2023, Leconte2013}. The N$_2$ + 376 ppm CO$_2$ composition was chosen as the reference case due to its established use in prior 3D simulations with the Generic-PCM model. O$_2$ is typically omitted because it is neither a greenhouse gas nor uniformly distributed throughout the atmosphere. Incorporating O$_2$ would require accounting for photochemical processes, which adds significant complexity to the model and is outside the scope of this study. This simplified composition serves as a scientifically robust baseline for {validation.} 

However, in general, we prioritize the study of a predominantly H$_2$ atmosphere, a scenario of great interest given the abundance of H$_2$ in the universe (in the context of galaxy formation, \citep{Rigopoulou_2009}). Despite its abundance, there is a scarcity of studies on the habitability of planets with H$_2$ atmospheres, and existing calculations are only carried out using 1D models~\citep{Pierrehumbert_2011, Ramirez_2017, Koll_2019}. Due to its collision-induced absorption (CIA) \citep{Frommhold1993}, H$_2$ is responsible for high infrared absorption, especially with surface pressures higher than 1 bar, and this can sufficiently warm a planet to sustain surface liquid water far from its host star \citep{Pierrehumbert_2011, Ramirez_2017}. Such atmospheres, characterized by their low mass and substantial scale height, are easier to characterize using state-of-the-art telescopes, such as the James Webb Space Telescope (JWST,~\citep{Beichman, Lin}) and the upcoming Atmospheric Remote-sensing Infrared Exoplanet Large-survey (ARIEL) \citep{Ariel}. 

{A} low-pressure primordial H$_2$ atmosphere may have a relatively short lifespan on small planets due to escape mechanisms \citep{Owen_2013}. For example, Pierrehumbert and Gaidos (2011) \citep{Pierrehumbert_2011} demonstrated that a planet at 1 AU could lose over 100 bars of H$_2$ in 4.5 Gyr, making the 1-bar scenario transient without significant replenishment. However, studies indicate that such atmospheres can persist long enough for early biospheres to develop, particularly if supported by sustained volcanic outgassing \citep{Wordsworth_2013, Ramirez_2014, Ramirez_2017}. This highlights that while the 1-bar scenario simulated here may not represent stable configurations over geological timescales, it still provides valuable insights into the range of conditions under which habitability could emerge.

{Additionally,} increased surface pressures of H$_2$ can extend the atmospheric lifetime~\citep{MolLous}, allowing sufficient time for potential biosignatures to develop. These transient scenarios are designed to serve as boundary cases for exploring the interplay among atmospheric composition, surface conditions, and microbial adaptability, even if such configurations are unlikely to persist indefinitely in natural planetary settings. By investigating these limit cases, we aim to better understand how atmospheric dynamics shape the habitability potential of exoplanets.

Finally, we also focus on CO$_2$-dominated atmospheres. While CO$_2$ is a very common gas in the solar system, it has also been found in many gas-giant atmospheres, like WASP-39b~\citep{Alderson2022} using the JWST, and it is expected to be found in the secondary atmospheres of terrestrial planets \citep{Kaltenegger_2009, Deming2009, Lustig_Yaeger_2019}. CO$_2$ levels on terrestrial planets are regulated over geologic timescales by the carbonate-silicate cycle, which balances volcanic outgassing of CO$_2$ with its removal via weathering into carbonates \citep{Kasting2019}. This cycle is a key thermostat mechanism, stabilizing surface temperatures conducive to liquid water, provided there are active surface renewal mechanisms, such as plate tectonics or volcanic activity, to recycle carbon between the atmosphere and lithosphere. While we do not explicitly model these long-term processes, our simulations assume that such mechanisms could establish and sustain the atmospheric scenarios we investigate. Our climate model simulates climate conditions over years or decades, whereas the carbonate-silicate cycle and other regulatory mechanisms operate on geological~{timescales.}

On the one hand, CO$_2$ is a very potent greenhouse gas and can provide habitable conditions at greater orbital distances \citep{Kadoya_2019}, but on the other hand, it is only efficient below a certain maximum partial pressure. In fact, for high surface pressures, the increased albedo due to Rayleigh scattering counterbalances the increase in the greenhouse effect of CO$_2$~\citep{KASTING}.

In addition to determining habitable conditions and surface habitability, the composition of the atmosphere significantly influences the potential for life to survive, grow, and reproduce. It can provide essential elements, such as CO$_2$ and H$_2$, which can serve as key resources for various metabolic pathways: CO$_2$ provides carbon for photosynthetic and autotrophic organisms, while H$_2$ functions as an electron donor for many anaerobes, including methanogens and sulfur-reducing bacteria. Additionally, atmospheric components like ozone (O$_3$) shield microbial communities from harmful stellar radiation, protecting life in surface and near-surface environments \citep{Segura_2010}. Thus, the atmospheric composition is crucial to determine the metabolic and protective conditions necessary for microbial life. The atmospheric composition of the early Earth differed significantly from that of present-day Earth; therefore, the conditions for life to appear were distinct from those that support life on present-day Earth \citep{Kasting93, Mojzsis}. In particular, CO$_2$, CH$_4$, and water vapor are thought to have been prominent constituents, and the atmosphere was anaerobic \citep{Holland}. Evidence shows that the emergence of life occurred relatively soon after the formation of the Earth \citep{Javaux} under anaerobic conditions. Given this, we performed survival experiments with anaerobic N$_2$~+~CO$_2$ atmospheres as a representative of the early Earth. H$_2$- and CO$_2$-dominated atmospheres, as proposed by {Hu et al.} (2012) \cite{Hu_2012}, are also crucial compositions for exoplanet atmospheres. Additionally, CH$_4$-dominated atmospheres present an intriguing candidate for survival and growth experiments, given their abundance on some planets and moons in our solar system \citep{Kuiper1944, Janson_2013, Benneke_2019, Woitke_2021} and recent discoveries of exoplanets with significant CH$_4$ abundances such as K2-18 b and WASP-80 b \citep{Madhusudhan2023, Taylor2023}. {Due to challenges} in preparing the opacity table for a CO$_2$ + CH$_4$ gas mixture, the simulations were not performed for this type of atmosphere. However, the opacity tables in the form of correlated-k tables are currently being built for future simulations. 

Until recently, little effort has been made to estimate the survivability of Earth life under different atmospheres \citep{Balch_1979, Seager_2020}. While many exoplanets in the HZ orbit small red M-dwarf stars, such as TRAPPIST-1 and Proxima-b, these stars present unique challenges for life due to their high UV flux, which could negatively impact planetary atmospheres and habitability \citep{Segura_2010}. Additionally, it is not clear whether these planets could have retained their atmospheres altogether, as recent JWST observations seem to show \citep{Greene, Zieba_2023}. Although M-dwarfs have been a primary focus for habitability studies due to instrumental limitations, upcoming missions like PLATO \citep{Rauer2014} will target planets around solar-type stars. These missions, along with LIFE (Large Interferometer For Exoplanets) and HWO (Habitable Worlds Observatory), will focus on characterizing these planets to detect biosignatures and explore Earth-like habitability in more stable environments \citep{Damiano2023, Alei_2024}. Therefore, our study primarily investigates habitability around a Sun-like star, using 3D GCMs. And because these atmospheric compositions have not previously been investigated in 3D, we focus here on atmospheres dominated by H$_2$ and CO$_2$. Finally, we evaluate microbial adaptability under these conditions.

Recently, {Seager et al.} (2020) \cite{Seager_2020} assessed the survivability of (non-photosynthetic) \textit{E. coli} bacteria for a few hours in a hydrogen atmosphere. However, no astrobiological experiments have investigated the differences in growth patterns—such as the duration and intensity of the lag, log, and stationary phases of microbial growth in batch culture— across various atmospheric compositions, particularly in conditions where organisms need time to acclimate. By directly comparing cell densities under different atmospheres to those on present-day Earth, we can test the adaptability of bacterial life to various environmental conditions. We are specifically interested in how growth patterns differ between aerobic and anaerobic conditions, as well as under different anaerobic environments. To ensure valid comparisons across atmospheres, we selected \textit{E. coli}, a facultative anaerobe capable of surviving in both aerobic and anaerobic conditions. 

The aims of this research are therefore twofold. First, we aim to replicate habitable conditions found on terrestrial planets with varying atmospheric compositions and to define the HZ limits using a 3D GCM, with a particular focus on atmospheres dominated by H$_2$ and CO$_2$. Subsequently, we investigate the survival and growth rates of \textit{E. coli} K-12 under the surface conditions we find in the models for the different exoplanetary atmospheres. We compare our results to the atmosphere of present-day Earth through long-term (30 days) experiments. We analyze and compare growth patterns in both short-term and long-term exposure to different atmospheres.  Inspired by the work of Seager et al. (2020) \cite{Seager_2020}, we evaluate \textit{E. coli}'s adaptability and survival capabilities under different atmospheric compositions. In particular, and as for the climate simulations, we consider atmospheres of 100\% CO$_2$, 100\% H$_2$, and Earth's current atmosphere. We also investigate an anaerobic nitrogen-rich atmosphere consisting of 90\% N$_2$ + 10\% CO$_2$ and a methane-rich atmosphere consisting of 80\% CH$_4$ + 15\% N$_2$ + 5\% CO$_2$. In Section \ref{Materials and Methods}, we detail the GCM and experimental setup, followed by the presentation of the results in Section \ref{Results} and the conclusions in Section \ref{Discussion}.

%%%%%%%%%%%%%%%%%%%%%%%%%%%%%%%%%%%%%%%%%%
\section{{Materials and Methods} 
}\label{Materials and Methods}

By initially exploring habitable conditions for various atmospheric compositions with different surface pressures and, subsequently, defining the limits of the HZ, we establish the temperature range conducive to surface liquid water, which guides our choice of a realistic temperature setting for the subsequent survival experiments.

The methodologies employed in this study are outlined in the following subsections: we describe the GCM used for the simulations in Section~\ref{method_GCM}, followed by the sample preparation and setup for the \textit{E. coli} experiments in Section~\ref{method_life}.

\subsection{Simulations: Influence of the Atmosphere on the Limits of the HZ}\label{method_GCM}

{To study} the atmosphere's impact on the limits of the HZ, we used the Generic-Planetary Climate Model (G-PCM) ({\url{https://lmdz-forge.lmd.jussieu.fr/mediawiki/Planets/index.php/Overview_of_the_Generic_PCM}{, accessed on 8 November 2024),}} formerly known as the LMD Generic-GCM. The G-PCM is a GCM, and like all GCMs, it calculates the evolution of the 3D state of an atmosphere (temperature, wind patterns, clouds, and precipitation) based on its composition. This model has been used extensively to simulate many different planetary atmospheres in and out of our solar system  \citep{Charnay_2013, Forget_2013, Leconte_2013, Bolmont_2016, Turbet2016, Turbet_2018, Chaverot_2023}. 
{GCMs} are powerful tools designed to simulate planetary climates by solving complex equations governing fluid dynamics, radiative transfer, and thermodynamics in a 3D grid (longitude, latitude, and altitude) that covers the whole atmosphere. These models provide a realistic depiction of atmospheric behavior, including temperature distribution, wind patterns, cloud formation, and precipitation, over years or decades. Unlike simpler 1D models, GCMs capture spatial variations and feedback mechanisms, offering a more nuanced understanding of climate processes.

Using similar climate models, the inner limit has also been investigated for different rotations \citep{Way2020} and different host stars \citep{Kopparapu_2017} but with atmospheres similar to that of the Earth in terms of composition. In this context, we focus on an Earth-like planet (i.e., orbiting a Sun-like star) with an Earth-like rotation, and we specifically examine the effect of atmospheric compositions—H$_2$- and CO$_2$-dominated atmospheres—on the limits of the HZ, which have not yet been investigated using 3D models.

{The} inner boundary of the HZ can be defined either by the initiation of the runaway greenhouse effect \citep{KASTING} or the moist greenhouse effect \citep{Zsom_2013}. While some studies using other GCMs, such as CAM4 \citep{Wolf2015}, have suggested the onset of a moist greenhouse, we did not observe this effect in our simulations with the G-PCM \citep{Chaverot_2023}. The occurrence of the moist greenhouse limit remains a topic of debate, with differences observed between models. For instance, Kopparapu et al. (2017) \citep{Kopparapu_2017} discussed how atmospheric dynamics and feedback mechanisms may influence the onset of the moist greenhouse.

{In} this study, we assume that the inner limit of the HZ is determined by the initiation of the runaway greenhouse effect. This boundary corresponds to the orbital distance where the Simpson--Nakajima threshold is reached, triggering the runaway greenhouse effect \citep{Simpson, Nakajima, KASTING}. Previous studies \citep{Leconte_2013, Chaverot_2023} have shown that in 3D models, atmospheric dynamics can cause an overshoot of the Simpson--Nakajima limit by affecting the relative humidity. Therefore, we focus on 3D climate simulations, which provide a more nuanced understanding of atmospheric behavior near the inner edge of the HZ.

For each atmospheric composition we simulate, water is a variable component that can condense or evaporate as a function of the pressure/temperature conditions. Furthermore, for each composition, in order to find the inner edge, we simulate planets at different orbital distances. Finally, we also vary the surface pressure of the planets{---}1~bar,\, 2.5~bar, and 5~bar. {While} a 1-bar H$_2$ atmosphere may be transient due to escape processes \citep{Pierrehumbert_2011}, and a 1-bar CO$_2$ atmosphere might require regulation through long-term geological cycles \citep{Kasting2019}, these scenarios serve as boundary cases to study their radiative effects and potential impact on surface habitability. Using 3D GCM simulations, this study provides a first look at how these atmospheric compositions influence the inner edge of the habitable zone, offering valuable insights into the theoretical limits of habitability under these extreme conditions.

All considered planetary and stellar characteristics are shown in Table \ref{tab:char_phy}. The crucial model parameters are provided in Table S1 of the Supplementary Materials.

% Table 1
\begin{table}[H] 
\caption{Planetary and stellar parameters used for the simulations.}\label{tab:char_phy}
\newcolumntype{Y}{>{\centering\arraybackslash}X}
\begin{tabularx}{\textwidth}{CCC}
\toprule
\textbf{Planetary and Stellar Characteristics} & \textbf{Values}\\
\midrule
Radius and mass & Earth-like  \\
Eccentricity & 0  \\
Obliquity & 23.44$^{\circ}$ \\ 
Rotation period & 24 h  \\
Atmospheric Compositions & Earth-like or H$_2$- or CO$_2$-dominated \\
Surface pressures &  1 bar, 2.5 bar, or 5 bar \\
Stellar spectrum & Sun \\
\bottomrule
\end{tabularx}
\end{table}

{The} calculation of radiative transfer was performed using the correlated-k method, a widely used technique that efficiently determines net radiative fluxes in the atmosphere. At the core of this method is the use of opacity tables, which contain pre-calculated data on how atmospheric gases absorb and emit radiation across different spectral bands. These tables simplify the complex line-by-line radiative transfer calculations by grouping spectral lines into broader bands, such as infrared (IR) and visible (VI), and assigning average absorption coefficients for each band \citep{Liou, Lacis, Fu, Eymet_2016}.

{The} correlated-k tables used in our model are a specific format of these opacity tables, optimized for computational efficiency. They enable accurate and rapid computation of atmospheric radiation while maintaining fidelity to the physical processes involved \citep{Goody, Leconte2021, TAKAHASHI}. For each atmospheric composition, distinct correlated-k tables are required (refer to Table~\ref{tab:simulations}).  {For the} N$_2$ + 376ppm CO$_2$ atmosphere, we used the correlated-k table from Leconte et al. (2013) \citep{Leconte2013}. For the H$_2$- and CO$_2$-dominated atmospheres, new correlated-k tables were constructed, which can be found in
\citep{corrk_H2,corrk_CO2}. 
Additionally, the model accounts for other opacity contributions, including H$_2$-H$_2$ collision-induced absorption (CIA) and H$_2$-H$_2$ continuum absorption from the HITRAN {(High-Resolution Transmission Molecular Absorption)} database, which is a compilation of spectroscopic parameters used to calculate and simulate the transmission and emission of light in planetary atmospheres \citep{Gordon2017, KARMAN2019160}, and the MT-CKD3.3 {(Mlawer--Tobin--Clough--Kneizys--Davies)} water vapor continuum model \citep{MT-CKD}, which provides a parameterization of absorption processes caused by molecular interactions in the far wings of spectral lines and continuum regions (the MT-CKD3.3 version provides an updated and refined dataset for these processes) to account for absorption in the far wings of spectral lines and between-line regions. The CO$_2$ continuum absorption is excluded here, which is reasonable for HZ inner-edge calculations since water vapor dominates the continuum in these atmospheres.

% Table 2
\begin{table}[H]
\caption{Atmospheric compositions, correlated-k tables, and number of bands in the infrared and visible range (IR {$\times$}
 VI) used for the simulations in this work.}
\label{tab:simulations}
\begin{tabularx}{\textwidth}{m{5cm}<{\centering}m{5cm}<{\centering}C}
\toprule
\textbf{Atmospheric Composition} & \textbf{Correlated-k Table \textsuperscript{a}} & \textbf{IR {$\times$} VI Range} \\
\midrule
H$_2$ & Original correlated-k table \textsuperscript{b} & 40 $\times$ 35 \\
CO$_2$ & Original correlated-k table \textsuperscript{b}  & 40 $\times$ 35 \\
N$_2$ + 376 ppm CO$_2$ &  From Leconte et al. (2013) \cite{Leconte2013} \textsuperscript{c} & 38 $\times$ 36  \\
\bottomrule
\end{tabularx}
\noindent{\footnotesize{\textsuperscript{a} All tables encompass variable water vapor mixing ratios, and pre-defined temperature and pressure {grids.}  
\textsuperscript{b} Correlated-k table built for this study by incorporating absorption data files from the HITRAN database \citep{Gordon2017} and using the exo\_k library \cite{Exorem}{.}
\textsuperscript{c} Correlated-k table sourced from Leconte et al. (2013) \cite{Leconte2013}.}}
\end{table} 
%\subsection{Experiment: Atmospheric influence on survival and growth of life}\label{method_life}
\subsection{Experiment: Atmospheric Influence on Survival and Growth of \textit{E. coli}}\label{method_life}

In our experiments, we used different gases: (i) the present-day Earth atmosphere (78.08~\%~N$_2$, 20.95~\%~O$_2$, 0.93~\%~Ar, 420~ppm CO$_2$), referred to here as ``standard air''; (ii) pure CO$_2$; and (iii) pure H$_2$. These are the compositions we used in the climate simulations (see Table~\ref{tab:simulations}). We also explored two other mixtures: (iv) anaerobic N$_2$-rich (90~\%~N$_2$ + 10~\%~CO$_2$) and (v) CH$_4$-rich (80~\%~CH$_4$ + 15~\%~N$_2$ + 5~\%~CO$_2$).

\subsubsection{Sample Preparation}\label{Sample preparation}

We closely monitored microorganism growth over 30 days in the atmospheric compositions listed above—the longest study of its kind to date—using a method similar to that of Seager et al. (2020) \cite{Seager_2020}. We therefore extended the pioneering study of Seager et al., as their experiments ran for a much shorter period (8 hours) and only for initially pure H$_2$ atmosphere. As a model microorganism, we used the bacterium \textit{E. coli} K-12 CL83, obtained from the German Collection of Microorganisms and Cell Cultures GmbH \cite{DSMZ}. Our main argument for longer incubation is to capture any extended lag phases at the onset of bacterial growth under new atmospheric conditions, as well as to observe what happens after prolonged stationary-phase conditions (evidence of senescence).

The first set of containers, namely (i) standard air, (ii) pure~CO$_2$, (iii) N$_2$-rich, and (iv) CH$_4$-rich, were inoculated with 200~$\upmu$l of the starter culture of \textit{ E. coli} K-12 with an average cell count of 3.71 $\times$ 10$^7$~cells/ml, obtained from a CASY cell counter, and an average optical density value of OD$_{600}$ = 0.077. These containers were then placed in a shaker operating at 100 rpm (revolutions per minute) and maintained at a temperature of 22~$^{\circ}$C. This is a lower temperature than the ideal growth temperature of \textit{E. coli} \citep{Tuttle}; however, it falls in the surface temperature range of all simulated habitable planets based on the performed simulations (see Section \ref{Result_atm}). 

Due to the difficulties and dangers that come with dealing with pure hydrogen, we started the second set of bottles slightly after the first set. They contained pure H$_2$ (v) and were inoculated with 200~$\upmu$l of starter culture with a cell count of 4.214 $\times$ 10$^7$~cells/ml and an OD$_{600}$ value of 0.080. Subsequently, these bottles were placed in the same shaker at the same temperature.

\subsubsection{Experimental Setup}\label{Experimental setup}

\begin{figure}[H]
%\centering
\includegraphics[width=0.8\textwidth]
{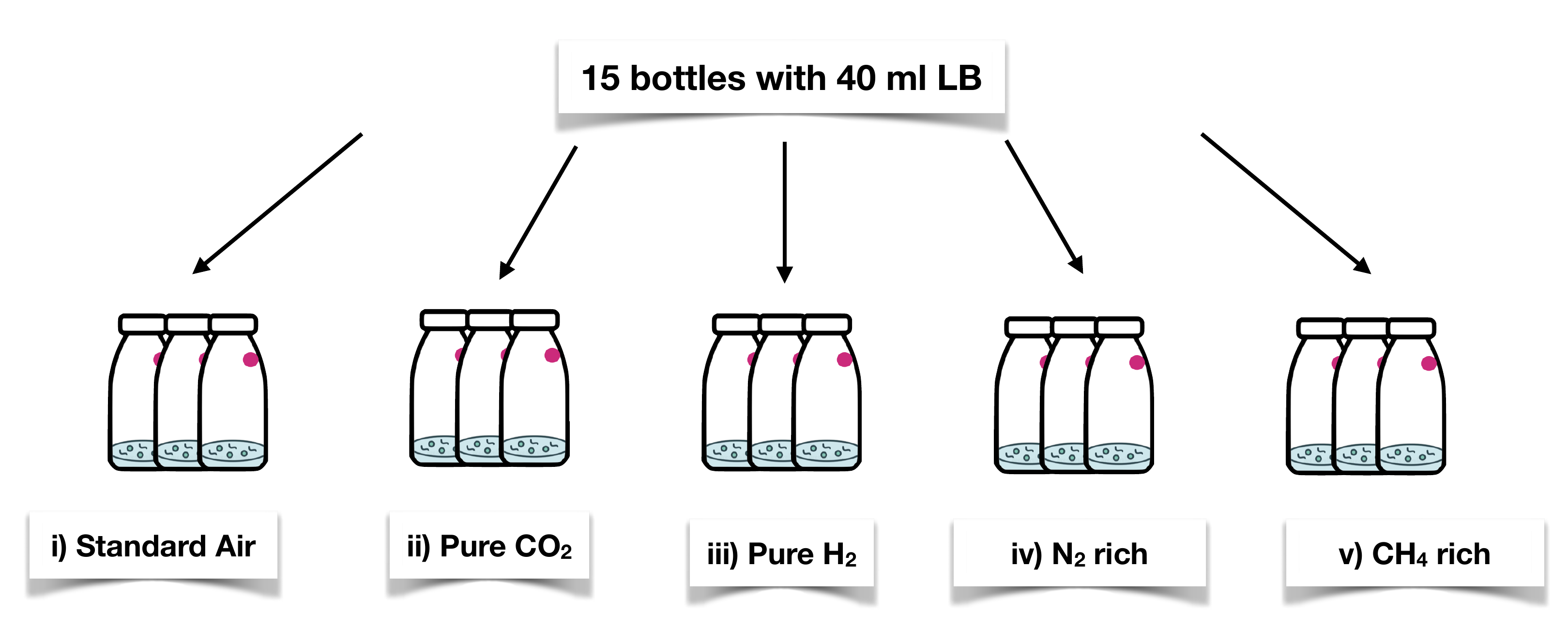}
\caption[Schematic of 15 bottles]{\label{fig:Bottles schematic}{Schematic} of the 15 bottles prepared for the laboratory experiments for the five distinct atmospheric compositions.} 
\end{figure}

A total of fifteen 250~ml borosilicate bottles were used, 3 per composition. Each bottle contained 40~ml of Lysogeny broth (LB) as a source of nutrients (refer to the schematic provided in Figure~\ref{fig:Bottles schematic}). To generate the intended environments in the bottles, we filled the airspace with the respective gases after removing air, i.e., oxygen, using the nitrogen flushing technique. A filter with a pore size of 0.2~$\upmu$m was used during the purging and filling process to prevent microbial contamination (see Figure S1 in Supplementary Materials). Before beginning the experiment, all containers, silicone rubbers, LB, and rocks were autoclaved at a temperature of 121~$^{\circ}$C. 

In addition to monitoring the growth of the bacteria, we also tracked oxygen levels to ensure gas concentration stability and verify anaerobic conditions within the bottles throughout the experiment. To do so, we used O$_2$ sensor dots from {PreSens }\cite{Presens}, affixed to the interior walls for noninvasive oxygen monitoring. 

%%%% 1st figure

%%%%%%%%%%%%%%%%%%%%%%%%%%%%%%%%%%%%%%%%%%
\section{Results}\label{Results}
\subsection{Inner Limit of HZ of Exoplanets with Distinct Atmospheres}\label{Result_atm}

As summarized in Table \ref{tab:simulations}, we modeled three different atmospheric compositions. By tracking the mean surface temperature and atmospheric water vapor content at three surface pressures (1 bar, 2.5 bar, and 5 bar) across varying orbital distances, we identified key planetary states, such as the runaway greenhouse state and the “habitable” state (defined here as one that hosts surface liquid water).

After successfully comparing our reference case (N$_2$~+~376~ppm CO$_2$) with the literature~\citep{Leconte2013, Chaverot_2023}, we performed simulations of H$_2$- and CO$_2$-dominated atmospheres. These results are presented in this section. The simulations were terminated at the onset of the runaway greenhouse state, where excessive water vapor accumulation in the atmosphere led to a sharp increase in surface temperatures.

\begin{figure}[H]
%\centering
\includegraphics[width=1\textwidth]
{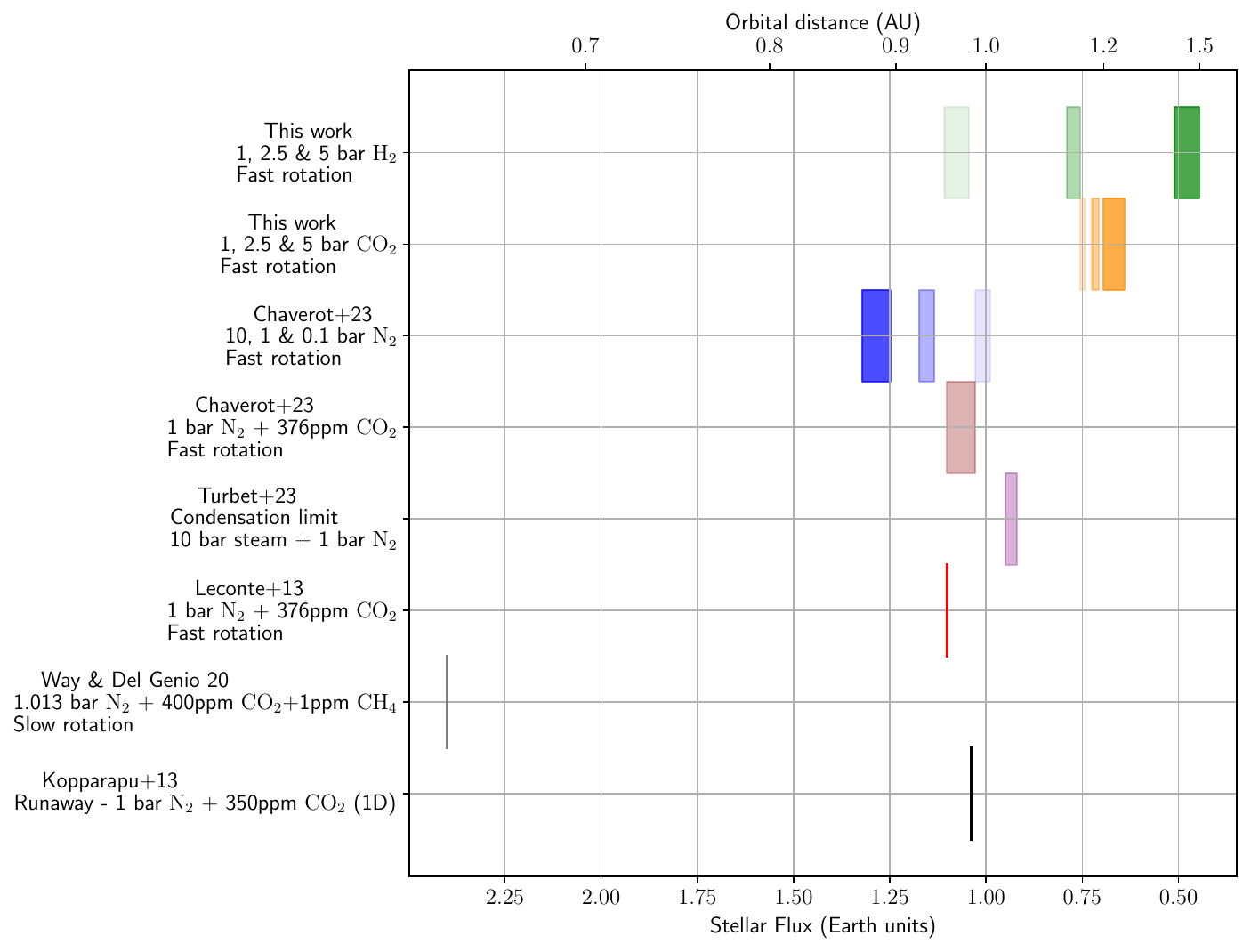} 
\caption[Inner HZ limits from different studies]{\label{fig:HZ_inner_limits}{Inner} limits of the HZ from various studies in 1D and 3D for N$_2$-, CO$_2$-, and H$_2$-dominated atmospheres. All limits \citep{Kopparapu, Leconte2013, Way2020, Chaverot_2023} except for one are computed for a ``cold'' start (i.e., water is initially condensed on the surface), while the limit of Turbet et al. (2023) \citep{Turbet_2023} is computed for a ``hot'' start  (i.e., water is initially steam in the atmosphere). For the 3 rows at the top, several pressures of the main background gas are shown in different shades: the lightest color corresponds to the lowest pressure, and the brightest color corresponds to the highest pressure. Our contributions to this plot are the first two rows: H$_2$- and CO$_2$-dominated atmospheres at different surface pressures (1~bar, 2.5~bar, and 5~bar).}
\end{figure}

Figure \ref{fig:HZ_inner_limits} shows the inner limit of the HZ for various atmospheric compositions and planetary rotations, computed by various codes (whether 1D or 3D) for Earth-sized planets around a Sun-like star. The contributions of this study to this image are the first two rows for CO$_2$- and H$_2$-dominated atmospheres at three different surface pressures (1~bar, 2.5~bar, and 5~bar). Among all the limits displayed in Figure~\ref{fig:HZ_inner_limits}, only one was actually computed using a 1D climate model. This limit is represented by a black line and was computed by Kopparapu et al. (2013) \citep{Kopparapu} for an Earth-like atmosphere. However, both Leconte et al. (2013) \citep{Leconte2013} (red line) and Chaverot et al. (2023) \citep{Chaverot_2023} (brown rectangle) have shown that this limit changes when computed using a 3D climate model. This illustrates the importance of accounting for the dynamics in the atmosphere, as it modifies relative humidity and, therefore, thermal emission. Additionally, Way and Del Genio (2020) \citep{Way2020} showed the importance of the rotation of the planet on the inner edge: a slow-rotating planet can remain habitable much closer to the host star than a fast-rotating planet (as shown in \citep{Leconte2013,Chaverot_2023}). Finally, we present the recent limit calculated by Turbet et al. (2023) \citep{Turbet_2023}, which represents the runaway condensation limit rather than the runaway greenhouse limit. This limit marks the point at which a steam atmosphere begins to condense on the surface. For a planet to sustain surface liquid water, it must, at some point in its lifetime, be positioned to the right of this limit. 

Our study contributes to this body of work by adding the inner HZ limits for H$_2$-dominated atmospheres at surface pressures of 1~bar, 2.5~bar, and 5~bar, represented by shaded green rectangles (with shading indicating pressure: the more transparent corresponding to the lowest pressure). Additionally, we provide the inner HZ limits for CO$_2$-dominated atmospheres at the same surface pressures (1~bar, 2.5~bar, and 5~bar), represented by shaded orange rectangles. As shown in the plot, N$_2$-dominated atmospheres exhibit different behavior with increasing surface pressure compared to H$_2$ and CO$_2$ atmospheres. N$_2$ itself is not a greenhouse gas, and in the case of an H$_2$O + N$_2$ atmosphere, as demonstrated by Chaverot et al. (2023) \citep{Chaverot_2023}, the presence of N$_2$ reduces the broadening of H$_2$O absorption lines, weakening the greenhouse effect compared to a pure H$_2$O atmosphere.  Consequently, lower surface pressures in N$_2$ atmospheres result in more effective warming than higher pressures. 

In contrast, H$_2$ and CO$_2$ are both strong greenhouse gases, and increasing surface pressure in these atmospheres amplifies their warming effects. However, the mechanisms by which they absorb infrared radiation differ. H$_2$ absorbs infrared radiation through collision-induced absorption (CIA), whereas CO$_2$ absorbs infrared radiation primarily via ro-vibrational transitions. Figure~\ref{fig:HZ_inner_limits} illustrates how the position of the inner edge of the HZ shifts with the pressure for CO$_2$ and H$_2$ atmospheres. For the two lowest pressures we considered here (1~bar and even 2.5~bar), the effect of CIA in the H$_2$ atmosphere is lower than the greenhouse effect of CO$_2$.  This can be seen in Figure~\ref{fig:HZ_inner_limits}, where the inner edge for 1~bar and 2.5~bar is closer for the H$_2$ atmosphere than for the CO$_2$ atmosphere. However, for the highest pressure considered here (5~bar), this is reversed: high pressure considerably enhances CIA in the H$_2$ atmosphere, whereas the effect of the pressure is less important for the CO$_2$ atmosphere. Additionally, an H$_2$-dominated atmosphere is more extended (having a larger scale height) than a CO$_2$-dominated atmosphere due to the lower molecular weight of H$_2$. This affects radiative transfer, as a more extended atmosphere increases the optical path length for infrared radiation. Furthermore, the competition between the greenhouse effect and Rayleigh scattering also plays a role in shaping the inner edge, with Rayleigh scattering being more significant for CO$_2$ than H$_2$.

%%%% 2nd figure

These contrasting behaviors highlight the fact that there is not one HZ, but that atmospheric composition and pressure play a significant role. This is particularly relevant in the context of observations, where most studies assess the placement of a planet in the HZ based on 1D climate models for an Earth-like atmosphere \citep{Kopparapu}. On the one hand, a planet deemed ``habitable'' by these standards can be ``unhabitable'' if it has a pure CO$_2$ atmosphere. On the other hand, a planet deemed ``uninhabitable'' by these standards can be ``habitable'' if it has a high N$_2$ pressure (or a slow rotation).

\subsubsection{H\texorpdfstring{$_2$-}{}Dominated Atmosphere}

\begin{figure}[H]
\centering
\includegraphics[width=1\textwidth]
{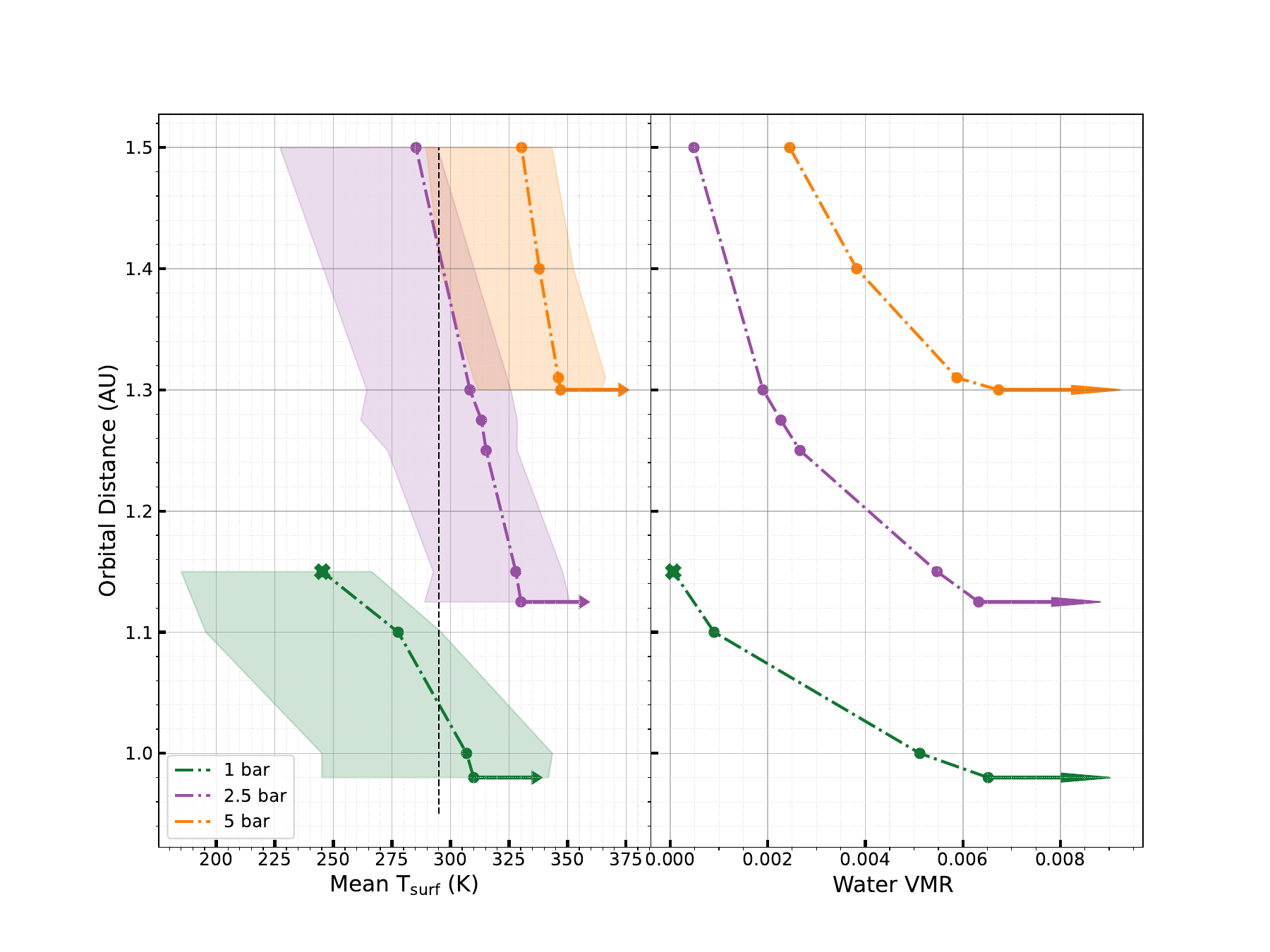}
\caption[Evaluation of tendencies for H$_2$ atmosphere]{\label{fig:Tsurf_H2}Dependence of the surface temperature (left panel) and water VMR (right panel) on the orbital distance of the planet for a H$_2$ atmosphere and different surface pressures (1 bar in blue, 2.5 bar in red, and 5 bar in green). The different filled circles correspond to steady states with surface liquid water (i.e., the habitable states). The arrows show the initiation of a runaway greenhouse state, while the crosses show snowball states. The dotted vertical black line in the left panel indicates 22 $^{\circ}$C, the chosen temperature for the laboratory experiments. We can see that it lies in the temperature range for all the habitable states. }
\end{figure}

Figure~\ref{fig:Tsurf_H2} illustrates the dependency of the mean surface temperature and the volume mixing ratio (VMR) of water on orbital distance for different pressures (1~bar, 2.5~bar, and 5~bar of dry air). We observe that an increase in surface pressure causes a gradual outward shift in the orbital distance at which the runaway greenhouse state initiates: 0.98~AU at 1~bar of dry air, 1.125~AU at 2.5~bar of dry air, and 1.3~AU at 5~bar of dry air. {Note that} achieving a fine interval of orbital distance is computationally expensive. For simulations of 15 planetary years, we require about one week of computation time on a cluster. Additionally, the mean surface temperature at the onset of the runaway greenhouse state rises with increasing surface pressures, reaching 310~K at 1~bar, 330~K at 2.5~bar, and 347~K at 5~bar of dry air. The warming effect of H$_2$ increases at higher pressures as the CIA effect increases as a factor of pressure and temperature as explained above. Therefore, at 1.5~AU with 2.5~bar and 5~bar, we still cannot reach the outer limit of the HZ for H$_2$ atmospheres. The right panel shows the change in the VMR of water as a function of orbital distance. The VMR is a critical parameter in atmospheric science, representing the proportion of a specific gas to the total gas mixture on a molar basis ($VMR=P_{H_{2}O}\  /\  P_{atm}$). 

Chaverot et al. (2022) \citep{Chaverot_2022} studied the behavior of a N$_2$ + H$_2$O atmosphere and how outgoing longwave radiation (OLR) evolves with a volume mixing ratio (VMR) for varying nitrogen surface pressures. Their study identified three key regimes—N$_2$-dominated, transitional, and H$_2$O-dominated—each characterized by changes in absorption line broadening (foreign vs. self-broadening) and shifts in the lapse rate. These regimes illustrate the radiative dynamics governing the Simpson--Nakajima limit and the critical transitions leading to runaway greenhouse effects. Using P-GCM, we confirmed the presence of these three regimes, demonstrating that they also emerge when global atmospheric dynamics are accounted for. This is a significant result, as the previous study by Chaverot et al. (2022)~\citep{Chaverot_2022} relied on a 1D model, which inherently excludes global atmospheric dynamics. The inclusion of 3D dynamics reveals an important difference: in our simulations, the threshold VMR required to initiate the runaway greenhouse state does not depend on the surface pressure of the dry air, whereas in 1D models, this dependency is present.

The right panel in Figure~\ref{fig:Tsurf_H2} shows the VMR of water for each surface pressure and orbital distance. A minimum VMR must be reached to trigger the runaway greenhouse state. In the case of H$_2$, an average VMR of $6.5 \times 10^{-3}$ is required to initiate the runaway state across all surface pressures. This threshold differs from the values reported in Chaverot et al. (2022) \citep{Chaverot_2022}, highlighting the impact of 3D atmospheric effects, such as large-scale circulation and global heat redistribution, which are absent in 1D models. This distinction underscores the importance of considering global dynamics in atmospheric modeling, particularly for exoplanet habitability studies. {The dataset} used for the plots can be found in ref. \cite{HZ_limits_H2}.

%%%% 3rd figure

\subsubsection{CO\texorpdfstring{$_2$-}{}Dominated Atmosphere}

\begin{figure}[H]
\centering
\includegraphics[width=1\textwidth]
{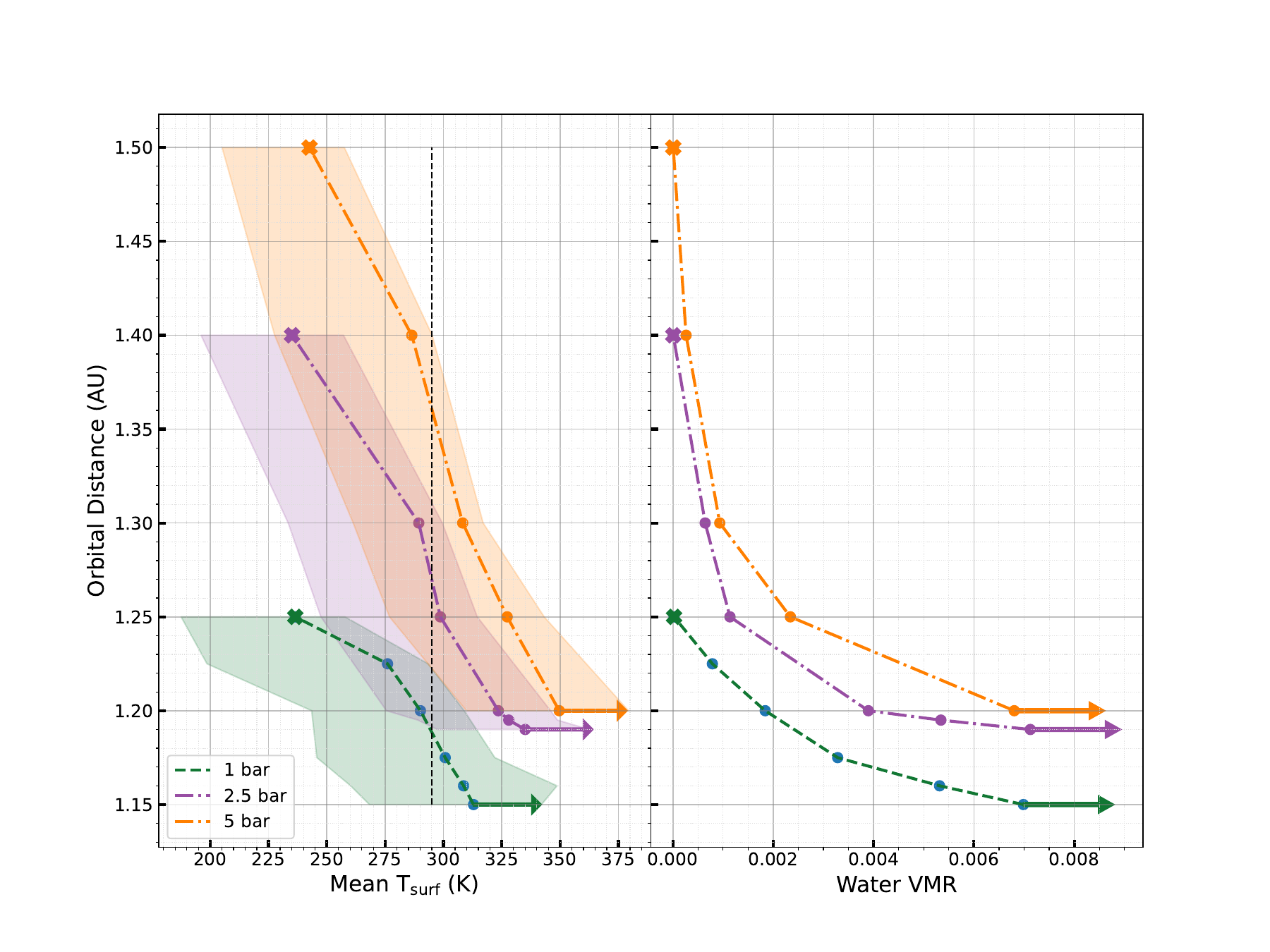}
\caption[Evaluation of tendencies for CO$_2$ atmosphere]{\label{fig:Tsurf_CO2}Dependence of the surface temperature (left panel) and water vapor VMR (right panel) on the orbital distance of the planet for a CO$_2$ atmosphere and different surface pressures (1 bar in blue, 2.5 bar in red, and 5 bar in green). The different filled circles correspond to steady states with surface liquid water (i.e., the habitable states). The arrows show the initiation of a runaway greenhouse state, while the crosses show snowball states. The dotted vertical black line in the left panel indicates 22 $^{\circ}$C, the chosen temperature for the laboratory experiments. We can see that it lies in the temperature range for all the habitable states.}
\end{figure}

Figure~\ref{fig:Tsurf_CO2} demonstrates that, similar to H$_2$ atmospheres, the onset of the runaway greenhouse state gradually shifts toward greater orbital distances with increasing surface pressure. The surface temperature at the onset of the runaway greenhouse state increases with rising surface pressures (313~K at 1~bar, 333~K at 2.5~bar, and 349~K at 5~bar), mirroring the behavior observed in H$_2$ atmospheres. However, we notice a smaller shift of the inner edge of the HZ (0.005~AU shift) when the surface pressure increases from 2.5~bar to 5~bar, in contrast to H$_2$ atmospheres, where the shift is relatively larger at the same surface pressures (0.185~AU shift). While the warming effect of H$_2$ intensifies with increasing surface pressure, the warming effect of CO$_2$ at high surface pressures is limited by Rayleigh scattering. In a CO$_2$ atmosphere, the maximum greenhouse limit is reached, and the planet becomes a snowball planet at 1.5~AU with a 5-bar surface pressure, whereas in an H$_2$ atmosphere, the planet is emerging from a runaway greenhouse state. The right panel illustrates that a minimum water VMR (an average of $6.9\times 10^{-3}$) must be achieved to trigger the runaway greenhouse state. The dataset used for the plots can be found in ref. \cite{HZ_limits_CO2}.

%%%% 4th figure

\subsection{Survivability and Growth of \textit{E. coli} in Distinct Atmospheres}

In this study, we monitored the growth of \textit{E. coli} under various atmospheric compositions. Based on the climate simulations, we opted for a temperature of 22~$^{\circ}$C (see Figures~\ref{fig:Tsurf_H2} and \ref{fig:Tsurf_CO2}). Considering that the ideal growth temperature for \textit{E. coli} is approximately 37~$^{\circ}$C, we anticipated a slower growth rate.

\begin{figure}[H]
%\centering
\includegraphics[width=0.95\textwidth]
{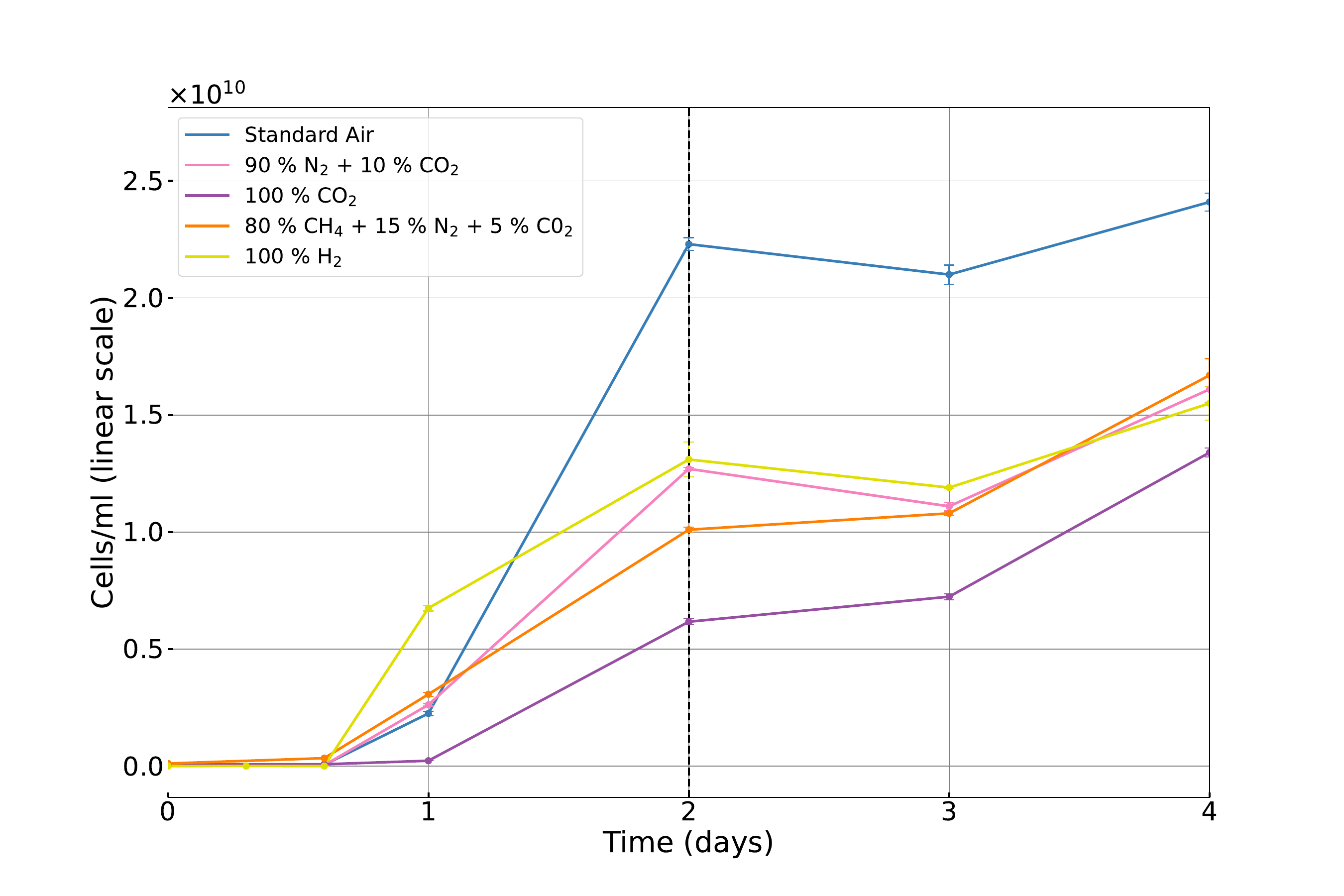}
\caption[Cells/ml for each atmosphere for the initial days]{\label{fig:CASY_short}{The} cell count of \textit{E. coli} K-12 cells/ml, plotted on a linear scale, for the initial days of each investigated atmospheric composition (standard air, 100 \% CO$_2$, 90 \% N$_2$ + 10 \% CO$_2$, 80\% CH$_4$ + 15\% N$_2$ + 5\% CO$_2$, and 100 \% H$_2$). The dotted black vertical line represents the end of the log phase. Error bars are shown in the figure but are very small and may not be clearly visible.}
\end{figure}

\begin{figure}[H]
\centering
\includegraphics[width=0.95\textwidth]
{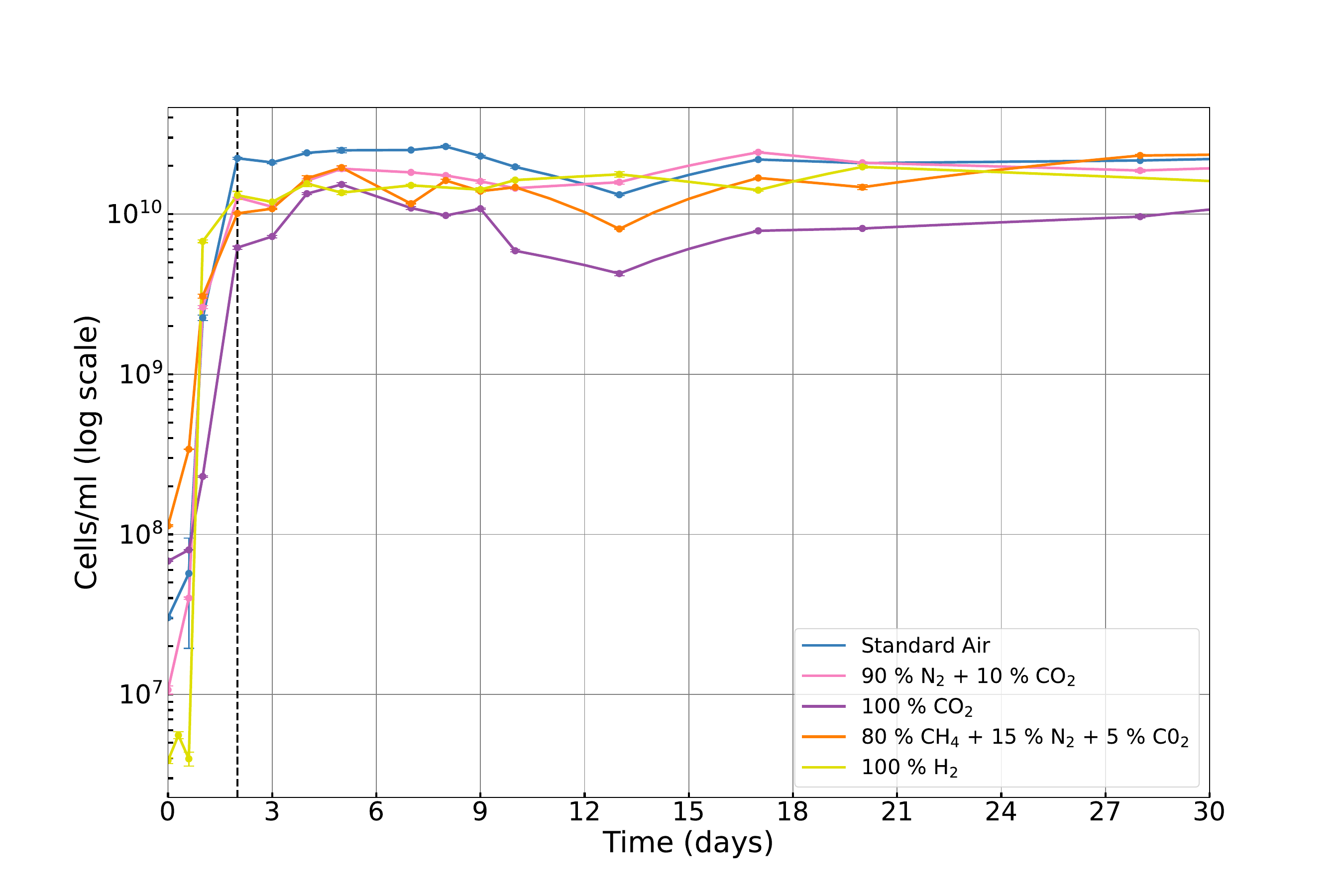}
\caption[Cells/ml for each atmosphere for 30 days]{\label{fig:CASY_long} The cell count of \textit{E. coli} K-12 cells/ml, plotted on a logarithmic scale over 30 days, focusing on the growth changes during the stationary phase. The end of the log phase is represented by the black dotted vertical line.}
\end{figure}

To analyze the growth dynamics, \textit{E. coli} cells were counted using the CASY cell counter at various intervals over a 30-day period. Given the nature of microbial growth, in order to effectively capture both short-term and long-term trends, the data were plotted on both linear and logarithmic scales (see Figures \ref{fig:CASY_short} and \ref{fig:CASY_long}). For each atmospheric condition, we measured the cell density using three independent biological replicates (three bottles per atmosphere) to account for variability among samples. The error bars in the figures represent the standard deviation of these measurements across the three replicates. 

Figure \ref{fig:CASY_short} illustrates the growth dynamics of \textit{E. coli} during the first 4 days under various atmospheric compositions. The data were plotted on a linear scale to provide a clear view of the initial cell proliferation rates. Given the rapid growth typically observed during this phase—assuming the absence of a prolonged lag phase—the linear scale enables a clear comparison of the absolute changes in cell counts over time. 

%%%% 5th Figure

On the day of inoculation, \textit{E. coli} entered the lag phase, during which it acclimatized to the new conditions, and no growth was observed. By the first day after inoculation, cell densities had increased in standard air, CH$_4$-rich, N$_2$-rich, and pure H$_2$ atmospheres. While cell densities increased similarly in standard air, CH$_4$-rich, and N$_2$-rich atmospheres, a slightly stronger increase was observed in the pure H$_2$ atmosphere. In contrast, cell densities remained lower in pure CO$_2$, indicating a longer acclimation period. Growth in standard air was the strongest by the second day, reflecting the suitability of oxygen-rich environments for aerobic respiration, the most energy-efficient metabolic pathway for many organisms. In the case of H$_2$, cell densities remained lower than those in standard air but were still higher than in the CH$_4$-rich and N$_2$-rich atmospheres. The rapid adaptation of \textit{E. coli} to pure H$_2$ suggests that hydrogen-rich atmospheres can support anaerobic microbial life once acclimatization occurs. Pure CO$_2$, however, consistently presented the most challenging environment, with significantly slower growth.

To assess long-term growth dynamics over the 30-day period, data were plotted on a logarithmic scale (see Figure \ref{fig:CASY_long}). This approach mitigates the impact of the fluctuations seen in the linear representation, particularly during the stationary and senescence phases, allowing for a clearer visualization of trends in cell survival and adaptation over extended periods. The logarithmic scale is especially useful for distinguishing differences in growth rates during phases where cell density changes over several orders of magnitude. Figure~\ref{fig:CASY_long} shows that after the first 3 days (represented by the dotted black vertical line), growth entered the stationary phase, where the rate of cell death equals the rate of cell division. The similar long-term growth rates observed in hydrogen, methane, and nitrogen-rich atmospheres suggest that \textit{E. coli} can adapt to these conditions over time.

This suggests that planets with anaerobic atmospheres dominated by H$_2$, CH$_4$, or N$_2$ may still be capable of supporting microbial life, even if the initial growth is slower than in standard air. The ability to adapt to less favorable conditions implies that life could persist on such planets, given sufficient time for acclimatization. In contrast, the cell density in pure CO$_2$ remained lower during the stationary phase, confirming it as the most challenging atmosphere for \textit{E. coli} growth. The consistently poor growth in pure CO$_2$ highlights the limitations of this gas in supporting life, particularly for heterotrophic organisms like \textit{E. coli}. While CO$_2$ can serve as a carbon source for some organisms (e.g., autotrophs), it lacks the properties necessary to sustain efficient metabolic processes in most heterotrophs. Planets with CO$_2$-dominated atmospheres may, therefore, require specialized life forms or adaptations, such as chemotrophs or extremophiles, to survive.

%%%% 6th Figure

%%%%%%%%%%%%%%%%%%%%%%%%%%%%%%%%%%%%%%%%%%
\section{Discussion}\label{Discussion}

This paper aimed to bridge the physical and biological factors that influence the habitability of exoplanets. One of our key objectives was to define the limits of the HZ for planets dominated by H$_2$ and CO$_2$ using 3D climate modeling, specifically the Generic-PCM~model.

Our results indicate that the warming effect of H$_2$ is particularly strong due to H$_2$-H$_2$ collision-induced absorption, pushing the inner edge of the HZ to further orbital distances than CO$_2$-dominated atmospheres. Specifically, we found that the inner edge for H$_2$ atmospheres could extend to 1.4 AU at 5 bar, while CO$_2$ atmospheres at the same pressure reached their inner limit at 1.2 AU. This demonstrates the profound impact of atmospheric composition on planetary climate and highlights how H$_2$ atmospheres can extend the habitable zone further from their host stars. In contrast, CO$_2$-dominated atmospheres exhibit limited warming at higher surface pressures due to Rayleigh scattering, which reduces the greenhouse effect. At higher pressures, planets with CO$_2$ atmospheres may struggle to maintain surface temperatures conducive to liquid water. Additionally, our results show that higher surface temperatures are required to initiate a runaway greenhouse effect at greater pressures (e.g., H$_2$: 347 K at 5 bar and CO$_2$: 349 K at 5 bar). {Although} some of the atmospheric scenarios presented here (1-bar H$_2$ and CO$_2$) are simplified and may not persist over geological timescales due to processes like hydrogen escape and carbonate-silicate cycling, they nonetheless provide valuable insights into the radiative effects of these gases on habitability. Understanding these transient configurations is critical for interpreting the potential habitability of exoplanets, particularly during specific evolutionary stages.

These insights illustrate the complexity of planetary atmospheres under varying conditions—factors such as atmospheric circulation, heat distribution, and other dynamics are better captured by 3D models, whereas 1D models fail to fully address these complexities. The use of 3D simulations enables us to more accurately assess the interaction between surface pressure, temperature, and atmospheric composition. This is especially important for exoplanetary habitability studies, as many previous models relied on 1D approaches, which are limited in their ability to predict climate behaviors across different atmospheres, especially by underestimating thermal emission for instance \citep{Leconte_2013}.

Together with this modeling approach, we investigated the growth of \textit{E. coli} under various atmospheric compositions, mirroring GCM setups and simulating potential exoplanetary environments. Our findings clearly showed that atmospheric composition significantly influences microbial growth. By the first day, \textit{E. coli} growth had increased in standard air, CH$_4$-rich, N$_2$-rich, and pure H$_2$ atmospheres, indicating that these environments support early exponential growth. The rapid adaptation of \textit{E. coli} to the H$_2$ atmosphere, with slightly stronger growth compared to the other atmospheres, was particularly noteworthy. In contrast, growth under pure CO$_2$ was slower, indicating a longer acclimation~period.

By the second day, growth in standard air was the strongest, reflecting the suitability of oxygen-rich environments for aerobic respiration. Although growth in pure H$_2$ remained lower than in standard air, it still surpassed that of CH$_4$-rich and N$_2$-rich conditions, suggesting that hydrogen-rich atmospheres can support anaerobic life once organisms have acclimated. Pure CO$_2$, however, presented the greatest challenge, with the lowest growth rates observed in both the short-term exponential phase and the long-term stationary phase.

Long-term analysis showed that \textit{E. coli} reached similar growth levels in standard air, CH$_4$-rich, H$_2$-rich, and N$_2$-rich atmospheres by the stationary phase, suggesting that microbial life can adapt to these environments over time. The consistently lower growth observed in pure CO$_2$ atmospheres suggests that CO$_2$-dominated environments may inhibit microbial survival and proliferation. This finding has direct implications for planets like Mars and Venus, which are believed to have had CO$_2$-rich atmospheres. While Venus currently has extreme surface conditions with a dense CO$_2$ atmosphere, it may have had a more temperate climate in the past, potentially supporting liquid water. Similarly, early Mars likely had a thicker atmosphere, as evidenced by ancient river valleys and lake beds, which could have created favorable conditions for life.

Our simulations align well with these experimental findings, offering insights into the habitability of exoplanets with different atmospheric compositions. For example, we know from both the lab and simulations that an N$_2$ + CO$_2$ atmosphere can support the survival of \textit{E. coli} while also sustaining surface liquid water. Similarly, our experiments demonstrated that \textit{E. coli} adapts well to hydrogen-dominated atmospheres. Hydrogen-rich planets, as shown by our simulations, are favorable for habitability due to H$_2$'s strong greenhouse effect, allowing surface liquid water to exist even at 1 bar. However, the tendency of thin H$_2$ atmospheres to escape from small terrestrial planets suggests that dense H$_2$ atmospheres are more suitable for sustaining both life and liquid water over long time scales.

In methane-rich atmospheres, \textit{E. coli} also adapted well over time, indicating that CH$_4$-dominated planets may be viable candidates for hosting life. While our current simulations did not focus on CH$_4$ atmospheres due to a lack of opacity data, it would be valuable to explore the limits of the habitable zone for methane-rich atmospheres, given their potential to support microbial life. In contrast, our lab results showed poor \textit{E. coli} growth under pure CO$_2$, suggesting that CO$_2$-rich atmospheres may require specialized organisms like extremophiles for life to thrive. This finding suggests that, although CO$_2$-dominated atmospheres are common in planetary systems, they pose challenges to microbial survival, limiting their habitability. Our simulations further support this by showing that a dense CO$_2$ atmosphere can eventually cool the planet due to Rayleigh scattering, restricting the ability to maintain liquid water at high surface pressures. Planets with CO$_2$ atmospheres below a certain pressure threshold, or those capable of supporting extremophiles, may be more suitable for habitability. Additionally, introducing other greenhouse gases like H$_2$ could mitigate the cooling effects of Rayleigh scattering and better support surface liquid water and life.

Overall, these results highlight both the resilience of \textit{E. coli} in adapting to diverse atmospheric conditions and the critical role atmospheric composition plays in determining microbial survival. The substantial growth inhibition observed under pure CO$_2$ underscores the importance of atmospheric composition for astrobiology and the search for life in extraterrestrial environments.

{Although} these findings are rooted in an Earth-centric framework, focusing on ``life as we know it'', the broader implications of habitability extend beyond these specific parameters. The concept of the HZ is limited by our understanding of life on Earth and should not be interpreted as excluding the possibility of habitability in non-water-based environments or for non-terrestrial life forms. As discussed by Seager (2013) \citep{Seager2013} and Bains et al. (2024) \citep{Bains2024}, life in entirely different atmospheric compositions could thrive under conditions vastly different from those studied here. Thus, our study highlights the importance of atmospheric composition and pressure for habitability while acknowledging the limitations of our Earth-centric perspective.

Our study not only enhances our understanding of the HZ and microbial adaptation but also demonstrates the value of a dual approach that combines climate simulations with laboratory survival experiments. This integrated methodology provides a promising framework for investigating habitability across different planetary settings, including early Earth, early Mars, and exoplanets. By exploring both atmospheric conditions and microbial survival, we gain a deeper understanding of the complex factors that influence habitability, paving the way for future research on the potential for life beyond our solar system.

\vspace{6pt}

\supplementary{The following supporting information can be downloaded at:  \linksupplementary{s1}, Table S1: Model parameters used for the simulations; Figure S1: Photo of a bottle used for the pure CO$_2$ atmosphere experiment (ii). Like this bottle, each bottle has two needles with a 0.2 µm sterile filter for flushing with nitrogen and consequently filling the head-space with the desired atmosphere.} 

%%%%%%%%%%%%%%%%%%%%%%%%%%%%%%%%%%%%%%%%%%
\authorcontributions{Conceptualization, A.K., E.B., B.W.I.,  and D.F.M.; Formal analysis, A.K.; Funding acquisition, E.B. and D.F.M.; Investigation, A.K. and J.Q.F.; Methodology, A.K.; Project administration, A.K.; Resources, E.B., B.W.I., and D.F.M.; Software, A.K.; Supervision, E.B. and D.F.M.; Validation, A.K., E.B., G.C., and Siddharth Bhatnagar; Visualization, A.K.; Writing---original draft, A.K.; Writing---review and editing, A.K., E.B., G.C., B.W.I., and S.B. {All authors have read and agreed to the published version of the manuscript.} 
} 

\funding{This work was carried out within the framework of the NCCR PlanetS, supported by the Swiss National Science Foundation under grants 51NF40\_182901 and 51NF40\_205606. A.K. and E.B. acknowledge the financial support of the SNSF (grant number: 200020\_215760). G.C. acknowledges the financial support of the SNSF (grant number: P500PT\_217840).}

\institutionalreview{Not applicable.}

\informedconsent{{Not applicable.}}

\dataavailability{All data used for this study, including the correlated-k tables prepared for H$_2$ and CO$_2$ atmospheres and the results of the simulations performed for both atmospheric compositions, can be found on Zenodo and in \citep{corrk_CO2, corrk_H2, HZ_limits_H2,HZ_limits_CO2}.}

\acknowledgments{The authors thank the Generic-PCM team for the teamwork development and improvement of the model. The computations were performed at the University of Geneva on the Baobab and Yggdrasil clusters. This research made use of NASA’s Astrophysics Data System. {We} acknowledge the use of ChatGPT-4 for grammar corrections and proofreading assistance, which contributed to the refinement of this manuscript. We confirm that all images and figures included in this manuscript are original and created by the authors. No copyrighted material has been used.} 

\conflictsofinterest{
The authors certify that they have no affiliations with or involvement in any organization or entity with any financial interest (such as honoraria; educational grants; participation in speakers’ bureaus; membership, employment, consultancies, stock ownership, or other equity interest; and expert testimony or patent-licensing arrangements) or non-financial interest (such as personal or professional relationships, affiliations, knowledge, or belief) in the subject matter or materials discussed in {this manuscript.}} 

%%%%%%%%%%%%%%%%%%%%%%%%%%%%%%%%%%%%%%%%%%
\begin{adjustwidth}{-\extralength}{0cm}
%\printendnotes[custom] % Un-comment to print a list of endnotes

\reftitle{References}

%=====================================
% References, variant A: external bibliography
%=====================================
%\bibliography{your_external_BibTeX_file}

%=====================================
% References, variant B: internal bibliography
%=====================================

%%%%%%%%%%%%%%%%%%%%%%%%%%%%%%%%%%%%%%%%%%
%% for journal Sci
%\reviewreports{\\
%Reviewer 1 comments and authors’ response\\
%Reviewer 2 comments and authors’ response\\
%Reviewer 3 comments and authors’ response
%}
%%%%%%%%%%%%%%%%%%%%%%%%%%%%%%%%%%%%%%%%%%
\PublishersNote{}
\end{adjustwidth}
\end{document}

\maketitle

\section*{This file includes:}
Supplementary tables for Generic-PCM setup and supplementary image for experimental setup.

\clearpage

% Table 1
\begin{table}[H] 
\caption{Model parameters used for the simulations.}\label{tab:char_model}
\begin{tabularx}{\textwidth}{Xl}
\toprule
\textbf{Model parameters}	& \textbf{Values}\\
\midrule
Resolution: Longitude x Latitude x Vertical Levels\textsuperscript{*} &60 x 48 x 30\\
Calculation of the dynamics (i.e. atmospheric transport) & Called every 96 s\textsuperscript{**} \\
Calculation of the physics (evaporation, condensation, etc.) & Called every 8 min\textsuperscript{**} \\
Radiative transfer & Called every 48 min\textsuperscript{**} \\
Ocean model & 2-layer ocean without dynamics \\
Ocean heat redistribution & false \\
\bottomrule
\end{tabularx}
\noindent{\footnotesize{\textsuperscript{*} \textbf{Resolution specifies the number of grid cells in each direction of the 3D grid used in the GCM.}}} \\
\noindent{\footnotesize{\textsuperscript{**} Numerical Time}} \\
\end{table}

%%%% 1st figure
\begin{figure}[H]
\centering
\includegraphics[width=0.7\textwidth]
{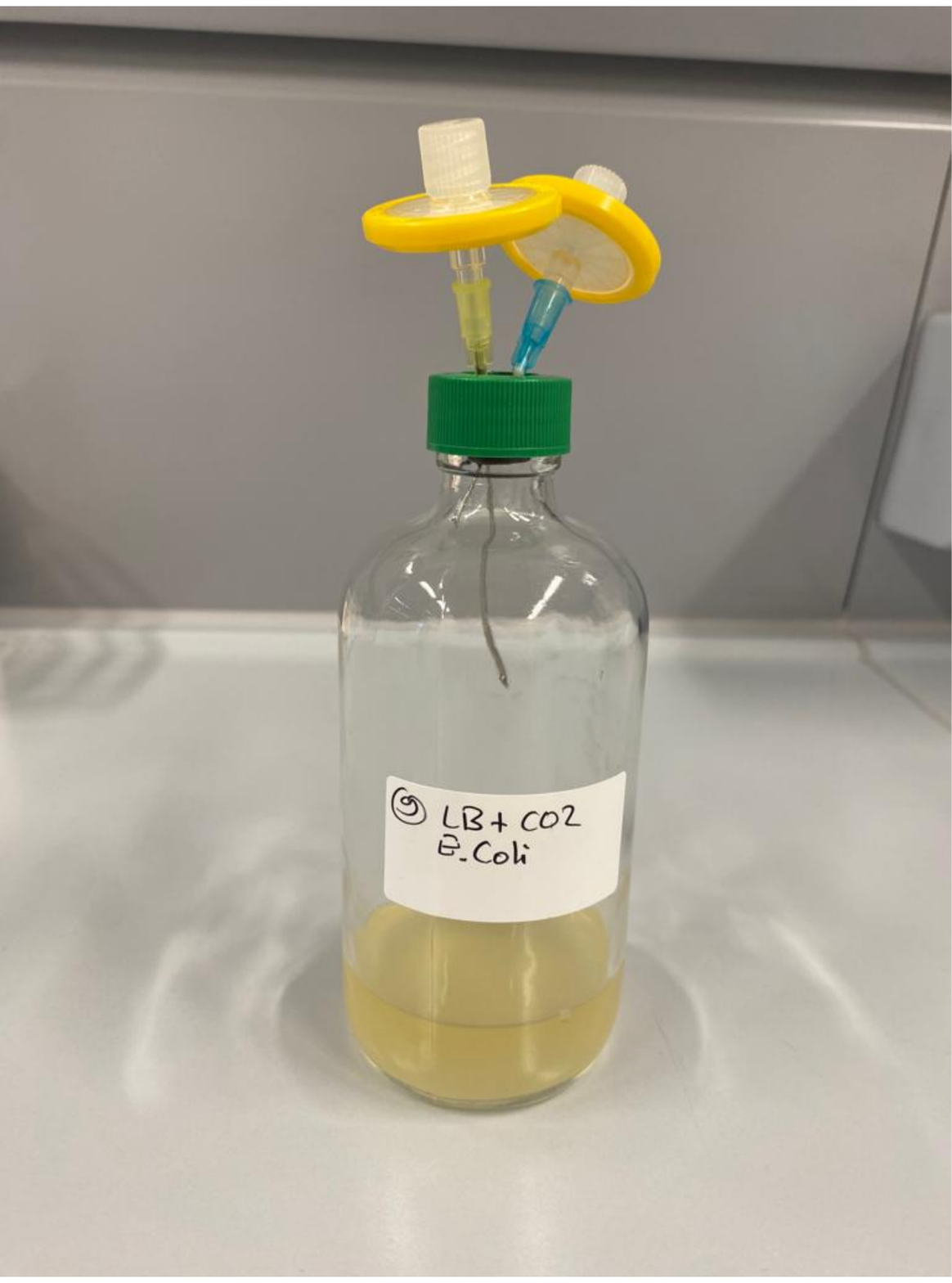}
\caption[Image of bottles used for the CO$_2$ atmosphere experiment]{\label{fig:bottles} Photo of a bottle used for the pure CO$_2$ atmosphere experiment (ii). Like this bottle, each bottle has two needles with a 0.2 µm sterile filter for flushing with nitrogen and consequently filling the head-space with the desired atmosphere.}
\end{figure}